\documentclass[conference]{IEEEtran}
\usepackage{amsmath}
\usepackage{amssymb,amsthm}
\usepackage[lined,boxed,commentsnumbered,ruled]{algorithm2e}
\usepackage{algorithmic} 
\usepackage{graphicx}
\usepackage{stmaryrd}
\SetSymbolFont{stmry}{bold}{U}{stmry}{m}{n}
\usepackage{multirow,url}
\usepackage{color, cite}
\usepackage{cite}
\usepackage{tikz,pgf}
\usetikzlibrary{fadings,shapes,arrows,shadows}
\usepackage{makecell}
\usepackage{array}
\usepackage{overpic} 
\usepackage{subfigure}
\usepackage{url}
\usepackage{epstopdf}
\usepackage{mathrsfs}
\usepackage{leftidx}
\usepackage{booktabs} 
\usepackage{float}
\usepackage{indentfirst}
\usepackage{latexsym,bm,amsmath,amssymb}
\usepackage{tabularx}
\usepackage{pifont}
\usepackage{threeparttable}
\usepackage[ruled]{algorithm2e}

\newtheorem{lemma}{Lemma}

\theoremstyle{plain}

\newcommand{\com}[1]{} 

\IEEEoverridecommandlockouts

\begin{document}

\title{Incentivizing Mobile Edge Caching and Sharing: An Evolutionary Game Approach }

\author{ 
Mingyu Li, Changkun Jiang, Lin Gao, Tong Wang, and Yufei Jiang 
\thanks{M.~Li,  T.~Wang, and Y.~Jiang are with the School of Electronics and Information Engineering, Harbin Institute of Technology, Shenzhen, China. 
C.~Jiang is with the Department of Computer Engineering, Shenzhen University, Shenzhen, China.
L.~Gao is with
the School of Electronics and Information Engineering, Harbin Institute of
Technology, Shenzhen, China, and the Shenzhen Institute of Artificial Intelligence
and Robotics for Society, Shenzhen, China. Email: gaol@hit.edu.cn.
(\emph{Corresponding Author: Lin Gao})~~~~
}
\thanks{This work is supported in part by the National Natural Science Foundation
of China (Grant No. 61972113), the Basic Research Project of Shenzhen
Science and Technology Program (Grant No. JCYJ20190806112215116,
JCYJ20180306171800589, and KQTD20190929172545139), and Guangdong
Science and Technology Planning Project under Grant 2018B030322004. This
work is also supported in part by the funding from Shenzhen Institute of
Artificial Intelligence and Robotics for Society.}
}

\maketitle

\addtolength{\abovedisplayskip}{-1mm}
\addtolength{\belowdisplayskip}{-1mm}

\begin{abstract}

\emph{Mobile Edge Caching} is a promising technique to enhance the content delivery quality and reduce the backhaul link congestion, by storing popular contents at the network edge or mobile devices (e.g. base stations and smartphones) that are proximate to content requesters.
In this work, we study a novel mobile edge caching framework, which enables mobile devices to \emph{cache} and \emph{share} popular contents with each other via device-to-device(D2D) links.
We are interested in the following incentive problem of mobile device users: \emph{whether} and \emph{which} users are willing to cache and share \emph{what} contents, taking the user mobility and cost/reward into consideration.
The problem is challenging in a large-scale network with a large number of users.
We introduce the evolutionary game theory, an effective tool for analyzing large-scale dynamic systems, to analyze the mobile users' content caching and sharing strategies.
Specifically, we first derive the users' best caching and sharing strategies, and then analyze how these best strategies change dynamically over time, based on which we further characterize the system equilibrium systematically.
Simulation results show that the proposed  caching scheme outperforms the existing schemes in terms of the total transmission cost and the cellular load.
In particular, in our simulation, the total transmission cost can be reduced by 42.5\%$\sim $55.2\% and the cellular load can be reduced by 21.5\%$\sim $56.4\%.


\end{abstract}

\IEEEpeerreviewmaketitle


\section{Introduction}

\subsection{Background and Motivations}

With the rapid development of the mobile Internet and the dramatic increase of mobile terminals, data services have shown an explosive growth in recent years. According to Cisco, the average monthly data usage worldwide will reach 2.9 billion TB by 2023, where video data will account for 82\% of all traffic \cite{Cisco-2020}. Moreover, researchers find that most of the data traffic is generated by the requests for a small number of popular contents, hence different contents' requests follow the Zipf's distribution \cite{Zipf-1999}. The repeated downloading of the same popular content by different mobile users will cause a large waste of backhaul resources and lead to a severe network congestion, significantly affecting the quality of service.

To solve such problems, researchers have proposed a novel scheme called \emph{edge caching} \cite{Yao-CST-2019}, which brings the function of content caching to the network edge. Edge caching preemptively transfers contents from the remote cloud at the core network to the edge network consisting of cellular base stations (BSs) or mobile user equipments (UEs), so that contents can be retrieved directly from the edge network when a user initiates a request \cite{Yao-CST-2019}. By caching popular contents closer to users, latency in getting content can be reduced. It can also avoid duplicate transmissions from cloud servers to end nodes, thus reducing the congestion of backhaul link.~~~~~~~~~~~~~~

In general, edge caching can be deployed in two kinds of edge devices: \emph{fixed devices} (e.g., BSs) and \emph{mobile devices} (e.g., UEs). There have been plentiful research focusing on caching on fixed devices at the network edge.
For example, \cite{Li-JSAC-2017} proposed a scheme of cooperative caching in software-defined networks, where each BS can obtain content from other BSs in the same macro cell, thereby reducing the backhaul traffic load.  \cite{Krolikowski-INFOCOM-2018} studied the joint optimization of edge nodes' caching placement and user-BS association to maximize the operator's utility. \cite{Shukla-JSAC-2018} optimized the duration of  contents' placement based on the collaborative caching at edge BSs to minimize the total cost of caching and downloading. Caching on fixed devices can effectively alleviate the network congestion and reduce the delay of content delivery. However, there are also many limitations, such as the relatively small device number and the limited serving range of each device.

To compensate for the limitations of fixed device caching, \emph{mobile edge caching} was  proposed to further reduce the network load, by caching contents on mobile UEs \cite{Li-JSAC-2018}.
{Today's mobile UEs generally have plenty of storage and computing resources to support content cache locally.
In addition, they can also act as mini caching servers and share the cached contents with each other via device-to-device(D2D) links, using licensed bands (e.g. LTE) or unlicensed bands (e.g. Bluetooth and Wi-Fi).}
Such a content caching and sharing scheme at mobile UEs is called \emph{crowdsourced caching} in \cite{Li-JSAC-2018}.
In \cite{Amer-TWC-2018}, Amer \emph{et al}. studied the optimal caching strategy in such a crowdsourced caching scheme to minimize the total energy consumption, where users' requests were served by the neighbors in the same cluster through single-hop D2D communication or served by any cluster in the same cell through inter-cluster collaboration.

Different from the fixed topology of fixed devices, the mobility of UEs leads to the \emph{non-fixed} topology, which makes the cache placement optimization in UEs very challenging. In particular, \cite{Song-TWC-2019} considered the user's mobility and modeled the arrival and departure of mobile users as a Poisson process.  \cite{Quer-TWC-2018} divided each cell into multiple sectors and users moved between cells, where user's mobility was modeled as a discrete Markov model.
Probabilistic caching is also a method suitable for caching in large-scale mobile networks. For example, \cite{Zhang-TVT-2018} placed contents in UEs according to the best cache probability to maximize the average hit rate.
However, these existing works studied the pure optimization problems, without considering the user incentive to cache and share contents.~~~~~~~~~~~




Another challenge of caching contents on mobile UEs is the \emph{user incentive}.
Specifically,
 users are often selfish in real world. That is, each user is more interested in caching his preferred contents, and hopes his neighbors to cache as many of his favorite contents as possible \cite{Chen-ICC-2016}.
 In addition, caching in UEs will occupy the storage space of devices, and the D2D transmission will also consume UEs' energy. Thus, an incentive mechanism is necessary to encourage the cooperative caching and sharing between UEs.
A few works have considered the incentive issue in edge caching by using game theory, which is a mathematical method for studying the interaction between incentive structures and the phenomenon of competition.
For example,  \cite{Sun-TVT-2016,Zheng-TWC-2018,Shen-GLOBECOM-2016} used game theory to analyze devices' caching strategy and equilibrium, but they only studied the game in a small network scenario with a limited number of users.
\cite{Jiang-ICC-2019} studied the game of caching strategy for a single content in a  large-scale edge network, but ignored the impact of user mobility and the diversity of contents.
In this work, we will consider the mobile edge caching in a large-scale network with an infinite number of mobile UEs, taking the incentive issue, together with the user mobility and content diversity, into consideration.

\subsection{Solution Approach and Contribution}

\begin{figure}
\vspace{-3mm}
 	\centering
\includegraphics[width=2.6in]{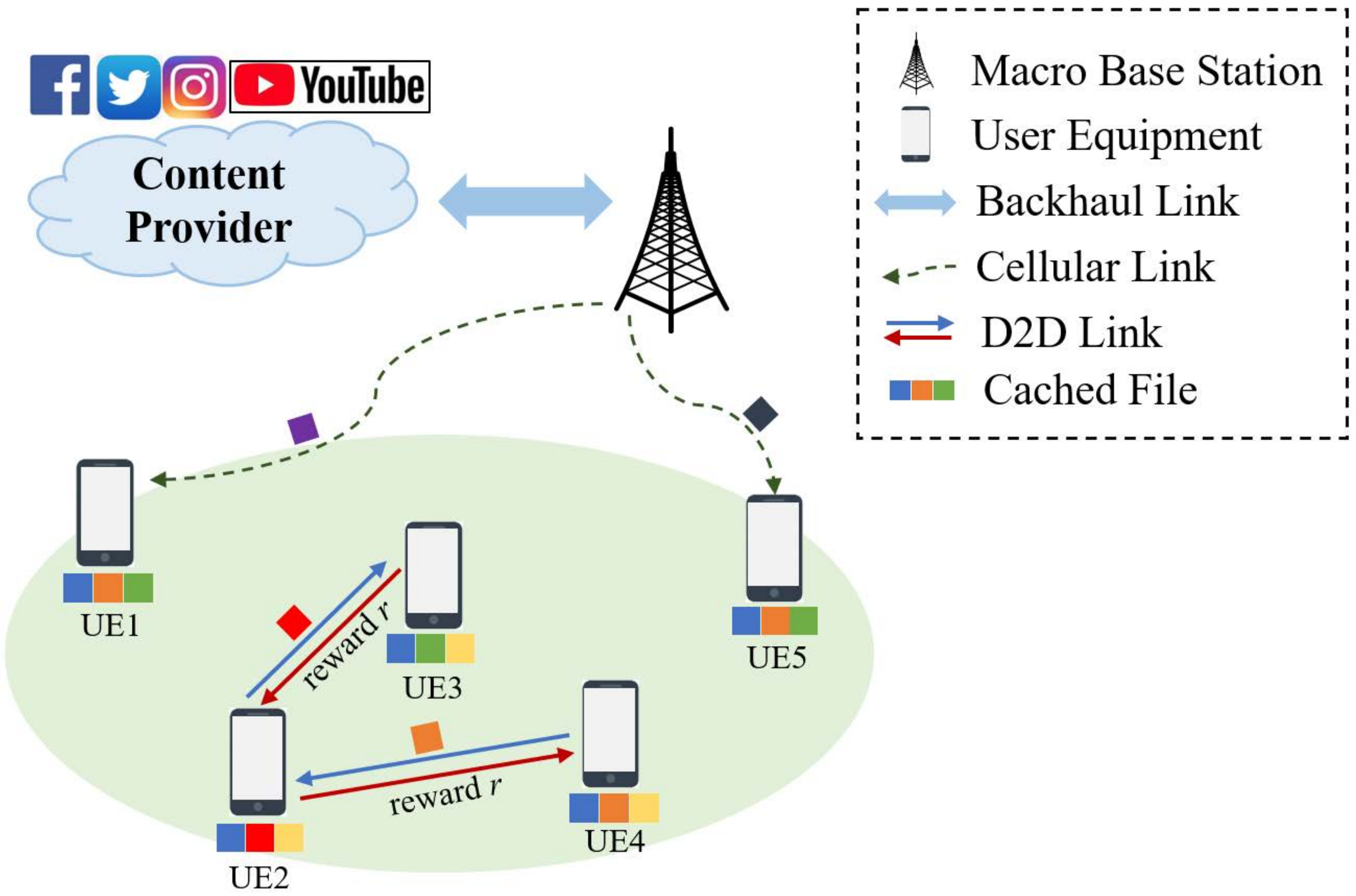}
\vspace{-3mm}
	\caption{Illustration of the Crowdsourced Caching Framework. }
\label{fig:fig-model}
\vspace{-5mm}
\end{figure}

Specifically, we study a crowdsourced caching framework for mobile edge caching, where contents are cached on mobile UEs. We consider a three-tier architecture, which consists of a content provider (CP), a macro base station (MBS), and a large number of mobile UEs.
UEs can cache contents (or files) in their device storages according to their respective cache strategies, and meanwhile share the cached contents with others via D2D links.
The content retrieval process is as follows.
When a UE initiates a content request, the requested content is first searched in the UE's local cache.
If the content is not found in the local cache, it is then searched in the caches of the requester's neighboring UEs.
The neighbor who caches the content becomes a mini server, and shares the content to the requester via the D2D link.
If the content is not found in all neighbors' cache, the request will be directed to the MBS and retrieved from the CP's remote cloud via the cellular link.
As content caching and sharing will introduce additional cost for UEs, the CP will provide certain monetary reward as incentive for UEs to cache and share contents.

Figure \ref{fig:fig-model} illustrates an example of such a  framework, where the squares with different colors represent different content files.
In this example, UE2 caches the blue, red, and yellow contents, and shares the red content with UE3 who does not cache the red content.
UE1 requests the purple content which is neither in its own cache, nor in its neighbors' caches, and thus obtain the content from CP's server through the MBS.

In such a crowdsourced framework, we are interested in the content caching and sharing problems of UEs, i.e.,
\begin{enumerate}
  \item Whether and which UEs are willing to cache and share contents, taking the user's mobility and incentive into consideration?
  \item What contents will be cached at which UE, considering the different popularity of different contents?
\end{enumerate}
The above problems are challenging due to the following reasons.
First, the strategies of different UEs are coupled with each other.
Second, the neighbors of each UE may change dynamically due to user's mobility.
Third, a practical network often consists of a large number of UEs.

We will analyze these problems by using the \emph{evolutionary game theory},  an effective tool for analyzing large-scale dynamic systems.
Specifically, we first derive the UEs' best caching and sharing strategies, and analyze their strategy dependence. Then we analyze how these best strategies evolve dynamically over time, based on which
we further characterize the system equilibrium systematically. 
Overall, the  key contributions of this work are summarized as follows:


\begin{itemize}
\item \emph{Novel Framework:}
We study a novel crowdsourced caching framework, which enables mobile UEs to cache and share popular contents locally, and thus can greatly expand the capability of mobile edge caching.

\item \emph{Novel Solution Technique:}
We introduce a novel solution technique--evolutionary game theory--to analyze the content caching and sharing problems of mobile UEs in a large-scale dynamic network.
By using this technique, we analyze the user strategy dependence and dynamics systematically, and characterize the game equilibrium.

\item \emph{Performance Evaluations and Insights:}
Simulation results show that our proposed framework outperforms the existing schemes in terms of both transmission cost and cellular load. Specifically, in our simulations, our proposed framework can reduce the total transmission cost by 21.5\%$\sim $56.4\%  and the cellular load by 24.1\%$\sim $57.8\%.

\end{itemize}

The rest of the paper is organized as follows. In Section \ref{section:model}, we present the system model. In Section \ref{section:game}, we analyze the equilibrium of the user-group strategies based on the evolutionary game. We present the simulation results in Section \ref{section:simulation}, and finally we conclude the paper in Section \ref{section:conclusion}.


\section{System Model}\label{section:model}

\subsection{Network Model}\label{section:model:network}

We consider a network of $U$ mobile UEs, each connecting to the Internet cloud through a MBS (as shown in Figure \ref{fig:fig-model}).
Let $ \mathcal{U} = \{1,\cdots,U\}$ denote the set of all UEs.\footnote{In this work, both ``user'' and ``UE'' refer to the mobile user device.} 
UEs can cache contents in their own storage and share with each other, hence form a crowdsourced caching network. 
When a UE initiates a content request, it first checks whether the content is available in its local cache; if not, then searches in the cache of nearby UEs via the D2D link; if the UE cannot get the requested content in all neighbors' cache, it connects to the cloud through MBS via the cellular link.
To encourage UEs to act as mini servers and share contents with others,  
the CP will provide a certain monetary reward as the incentive.

\subsection{Content Model}\label{section:model:content}

We consider a set of $F$ different content files, denoted by a set ${{\cal F}} = \{1,\cdots,f,\cdots,F\} $, where the size of file \emph{f} is $s_f$.  
Content popularity is the probability distribution of content requests from all users, namely the ratio of the number of requests for a certain content to all contents. The popularity of content  \emph{f} usually follows Zipf's distribution \cite{Li-X-JSAC-2018}, that is, the probability of any user requesting file \emph{f} is ${q_f}$:
\vspace{-1.5mm}
\begin{equation}
\vspace{-1mm}
{q_f} = \frac{{{{({R_f} + \varepsilon )}^{ - \beta }}}}{{\sum\limits_{i \in {{\cal F}}} {{{({R_i} + \varepsilon )}^{ - \beta }}} }},\forall f \in{{\cal F}}
\end{equation}where ${R_f}$ is the ranking of file \emph{f} in descending order of popularity in content library, $\varepsilon  \ge 0$ is the stationary factor, $\beta  > 0$ is the skew factor. When $\beta  = 0$, all files have the same popularity and users' requests will be evenly distributed on all content. When $\beta$ is a large number, the file with a lower index will have the higher popularity, which causes the users' requests to be concentrated on the first few popular files.

\subsection{User's Mobility}\label{section:model:mobility}
For mobile users in the network, D2D sharing is possible only when a pair of users are close enough to each other.
However, the neighbors of each UE may change dynamically due to user's mobility.
Based on the Erdos-R random graph theory, we propose an abstractive user's mobility model: users move randomly within a certain area, and the probability of each user encountering any other user is $\rho  \in [0,1]$.
The probability $\rho $ does not change over time, and the encounter events of each user are independent of each other.

Since we consider  a large-scale network scenario, we suppose that the number of mobile users in the network is infinite, namely $U \to \infty $. Therefore, the average number of neighbors that any user can encounter is $\psi = U \cdot \rho $.

\subsection{User Model}\label{section:model:user}

Due to the non-fixed topology of the mobile edge caching network, UEs adopt a probabilistic caching strategy: they first compute the probability of caching each content, and then select the set of contents to be cached based on the above probability distributions.
Denote ${x_{u,f}} \in [0,1]$ as the probability that UE \emph{u} caches file \emph{f}.

Because of the differences in UEs' performance, the cost of caching in different UEs varies.  We denote ${\alpha _u}$ as the cost of UE \emph{u} caching content in unit size and denote ${c_u}$ as the storage space of UE \emph{u}.

When a UE obtains contents from different objects, the transmission cost is various. Define the unit cost of a UE's request being satisfied locally as ${\omega _L}$, the unit cost of D2D sharing as ${\omega _D}$, and the unit cost of getting files via cellular links as ${\omega _B}$. Note that the three kinds of cost have a clear relationship of size, i.e., ${\omega _L}\ll{\omega _D}<{\omega _B}$. In addition, when a UE acts as a server to share contents via D2D links, the reward of unit transmission given to it by CP is \emph{r} (\emph{r} refers to the net value of the difference between the pure reward and the D2D output transmission cost). The total transmission cost and reward are directly proportional to the traffic.

The probability that UE \emph{u} requests file \emph{f} and its local cache can satisfy the request is ${x_{u,f}}{q_f}$, then the average local traffic load of user \emph{u} is

\vspace{-4mm}
\begin{equation}
T_u^L = \sum\limits_{f = 1}^F {{x_{u,f}}} {q_f}{s_f},\forall u \in {{\cal U}}
\vspace{-2mm}
\end{equation}

The probability that UE \emph{u} requests file \emph{f} but the local cache cannot satisfy the request is $(1 - {x_{u,f}}){q_f}$, then the average traffic load of UE \emph{u} being satisfied by D2D sharing is

\vspace{-3mm}
\begin{equation}
T_u^{in} = \sum\limits_{f = 1}^F {(1 - {x_{u,f}}){P_f}} {q_f}{s_f},\forall u \in{{\cal U}}
\vspace{-2mm}
\end{equation}where ${P_f}$ is the probability that any UE encounters at least one UE who has cached the file \emph{f} that it requests and the specific expression will be derived in Section  \ref{section:game:derivation}.

The probability of the situation that UE \emph{u} cannot find file \emph{f} in all neighbors' cache is $1 - {P_f}$. Therefore, the average traffic load of UE \emph{u} being satisfied via cellular link is

\vspace{-3mm}
\begin{equation}
T_u^B = \sum\limits_{f = 1}^F {(1 - {x_{u,f}})(1 - {P_f})} {q_f}{s_f},\forall u \in{{\cal U}}
\vspace{-2mm}
\end{equation}

The occupation of UEs' storage will affect their energy consumption and operating speed, thus bringing some cache cost.
Here we define the relationship between cache cost ${C_u}$ and cache size as a  convex function \cite{J-INFOCOM-2018}:

\vspace{-3mm}
\begin{equation}
{C_u} = \sum\limits_{f = 1}^F {{{({x_{u,f}}{s_f})}^2}{\alpha _u}} ,\forall u \in{{\cal U}}
\vspace{-2mm}
\end{equation}

When UE \emph{u} acts as a server to meet the requests of others, the reward of CP is proportional to the D2D output traffic:

\vspace{-2.5mm}
\begin{equation}
{V_u}{\rm{ = }}T_u^{out} \cdot r = \left(\sum\limits_{f = 1}^F {{x_{u,f}}{N_f}{s_f}} \right) \cdot r
\vspace{-2mm}
\end{equation}where ${{N_f}}$ refers to the average number of UEs connected to the UEs who have cached file  \emph{f}. The specific expression will be derived in Section \ref{section:game:derivation}.

When UEs perform D2D sharing in the edge network, their privacy cannot be fully guaranteed. Therefore, we introduce ${T_u^{out} \cdot \theta}$ to characterize the impact of users' privacy on the utility. The UE's utility consists of four parts: reward of D2D sharing, cache cost, transmission cost and price of privacy, and it is denoted by

\vspace{-3mm}
\begin{equation}
{U_u} = {V_u} - {C_u} - \underbrace {T_u^L \cdot {\omega _L} - T_u^{in} \cdot {\omega _D} - T_u^B \cdot {\omega _B}}_{{\rm{Transmission Cost}}}-T_u^{out} \cdot \theta
\vspace{-2mm}
\end{equation}

\subsection{Problem Formulation}\label{section:model:formulation}


In this work, the questions we ultimately want to study is: \emph{whether and which UEs are willing to cache and share contents, taking the user's mobility and incentive into
	consideration? what contents will be cached at which UE, considering the different popularity of different contents? }
We will analyze these problems by using the evolutionary game theory. Specifically, we first derive the UEs' best caching and sharing strategies, and analyze their strategy dependence. Then we analyze how these best strategies evolve dynamically over time, based on which we further characterize the system performance systematically.

\section{Game Equilibrium Analysis}\label{section:game}

In this section, we will first formulate the game problem of optimal caching and sharing strategies, and find the condition of equilibrium. Then we will derive some important variables in the model. Finally, we will find UEs' best responses of crowdsourced caching by using evolutionary game theory, and analyze how these best strategies change dynamically.

\subsection{Game Formulation}\label{section:game:formulation}

In the entire network, on the one hand, content caching and sharing will introduce additional cost for UEs. On the other hand, the CP will provide a certain monetary reward as the incentive for UEs to cache and share contents as mini servers.

Based on the crowdsourced caching framework, we consider mobile UEs as players in the game, and take the  transmission cost and incentive into account, in order to analyze the optimization of UEs' strategies. For each UE $u$, we establish an optimization problem with the goal of maximizing individual utility ${U_u}$, and its decision variable is the probability cache strategy of UE $u$:

\vspace{-2mm}
\begin{subequations}
	\vspace{-1mm}
	\begin{equation}\begin{aligned}\label{eq:sub1}  \mathop {\max }\limits_{{x_u}} ~~&{\rm{ }}{U_u}({x_u},{x_{ - u}})~~~~~~ \\
	\end{aligned}
	\vspace{-1mm}
	\end{equation}
	
	\vspace{-1mm}
	\begin{equation}
	\begin{aligned}
	\label{eq:sub2}
	s.t.~\sum\limits_{f = 1}^F {{x_{u,f}}{s_f} \le {C_u}} {\rm{   }} \end{aligned}
	\vspace{-1mm}
	\end{equation}
	
	\vspace{-0.5mm}
	\begin{equation}
	\begin{aligned}
	\label{eq:sub3}
	e \cdot T_u^{out} \le E
	\end{aligned}
	\vspace{-1mm}
	\end{equation}
	
	\vspace{-5mm}
	\begin{equation}
	~~~~~~~~~{x_{u,f}} \in [0,1],{\rm{  }}\forall f \in {{\cal F}}
	\vspace{-1mm}
	\end{equation}
\vspace{-3mm}
\end{subequations}

Here, ${x_u}=\{ {x_{u,1}}, \cdot  \cdot  \cdot ,{x_{u,f}}, \cdot  \cdot  \cdot ,{x_{u,F}}\} $. ${x_{ - u}}$ in \eqref{eq:sub1} denotes the cache strategies of all UEs except UE $u$. \eqref{eq:sub2} denotes the constraint of storage capacity. \eqref{eq:sub3} indicates the limit of power consumed by D2D sharing because of the actual situation that  the number of UEs served via D2D is limited. We define the energy consumption for transmitting data in unit size via D2D as $e$, then a UE's energy consumption of D2D transmission should be less than the threshold $E$.


It can be seen from \eqref{eq:sub1} that the utility function of UE $u$ is related to the cache strategies of other UEs, so the cache strategies of all UEs are coupled with each other. For the optimal solution $X = {\{ {x_u}\} _{U \times 1}}$, the final \emph{equilibrium} is reached if and only if the following conditions are satisfied:

	\vspace{-2mm}
\begin{equation}\label{eq:eqa9}
{U_u}(x_u^*,x_{ - u}^*) \ge {U_u}(x_u^{},x_{ - u}^*),\forall u \in {{\cal U}}
	\vspace{-1mm}
\end{equation}
The above equations imply that once an equilibrium is reached, none of the UEs has the incentive to change its strategy solely, and thus the system will keep on the equilibrium.


\subsection{Derivation of Important Variables}\label{section:game:derivation}

Before analyzing the optimization problem and game equilibrium, we first derive the specific expressions of ${P_f}$ and ${N_f}$, which are mentioned in Section \ref{section:model:user}.

\subsubsection{\textbf{Calculation of} ${P_f}$(Probability of encountering at least one UE caching file f)}

Consider any mobile UE \emph{u} in the network, and define the probability that another UE encountered by UE \emph{u} just has cached file \emph{f} as ${\eta _f}$, that is, the percentage of UEs caching file \emph{f} is

\vspace{-3mm}
\begin{equation}
{\eta _f} = \frac{1}{U}\sum\limits_{u = 1}^U {{x_{u,f}}} {\rm{, }}\forall f \in{{\cal F}}
\vspace{-2mm}
\end{equation}

Based on the definition in Section \ref{section:model:mobility}, the probability that UE \emph{u} meets any other UE is $\rho $, so the probability that UE \emph{u} encounters a fixed UE who has cached file \emph{f} is ${\eta _f}\rho$. Since there are a total of $U-1$ other UEs in the network, the probability that UE \emph{u} cannot encounter any UE who has cached file \emph{f} is ${(1 - {\eta _f}\rho )^{U - 1}}$. Therefore, the probability that UE \emph{u} meets at least one UE who has cached file \emph{f} is

\vspace{-3mm}
\begin{equation}
{P_f} = 1 - {(1 - {\eta _f}\rho )^{U - 1}}{\rm{, }}\forall f \in{{\cal F}}
\vspace{-1mm}
\end{equation}

Since $U$ grows to infinity, based on the principle of infinitesimal equivalence, ${P_f}$ can be further derived as

\vspace{-2mm}
\begin{equation}
\begin{aligned}
\begin{array}{l}
{P_f}=\mathop {\lim }\limits_{U \to \infty } 1 - {(1 - {\eta _f}\rho )^{U - 1}}\\
= \mathop {\lim }\limits_{U \to \infty } 1 - {(1 - {\eta _f}\frac{\psi }{U})^{U - 1}} = 1 - {e^{ - {\eta _f}\psi }}
\end{array}
\end{aligned}
\vspace{0mm}
\end{equation}

\subsubsection{\textbf{Calculation of} ${N_f}$(Average number of requesters connected to a UE caching file f)}

When UE \emph{u} requests file \emph{f} and meets multiple UEs who have cached file \emph{f} in the D2D communication range, he will randomly select one of these UEs for D2D sharing. Therefore, the probability that UE \emph{v} who has cached file \emph{f} is selected by requester \emph{u} is

\vspace{-2mm}
\begin{equation}
P_f^{{\rm{C}}} = \sum\limits_{k = 0}^{U - 2} {\frac{1}{{k + 1}} \cdot P_f^{(k)}} ,\forall f \in{{\cal F}}
\vspace{-2mm}
\end{equation}where $P_f^{(k)}$ denotes the probability that requester \emph{u} encounters $k$ UEs who also have cached file \emph{f} besides UE \emph{v}.

Except for UEs \emph{u} and \emph{v}, there are $U-2$ other UEs that requester \emph{u} may encounter. Therefore, the probability of requester \emph{u} encountering $k$ potential servers except UE \emph{v} obeys the binomial distribution:

\vspace{-3mm}
\begin{equation}
P_f^{(k)} = \left( \begin{array}{l}
U - 2\\
{\rm{   }}~~~k
\end{array} \right) \cdot {({\eta _f}\rho )^k} \cdot {(1 - {\eta _f}\rho )^{U - 2 - k}}
\vspace{-1mm}
\end{equation}
\vspace{-1mm}

Since $U$ is of a large value, ${u_f}\rho  = {u_f}\psi /U$ can be seen very small, so $P_f^{(k)}$ can be further derived as a Poisson distribution with arrival rate ${\eta _f}\psi $:

\vspace{-3mm}
\begin{equation}
\begin{aligned}
P_f^{(k)} &= \mathop {\lim }\limits_{U \to \infty } \left( \begin{array}{l}
U - 2\\
{\rm{   }}~~~k
\end{array} \right) \cdot {({\eta _f}\frac{\psi }{U})^k} \cdot {(1 - {\eta _f}\frac{\psi }{U})^{U - 2 - k}}\\
&= \frac{{{{({\eta _f}\psi )}^k}}}{{k!}} \cdot {e^{ - {\eta _f}\psi }}
\end{aligned}
\vspace{-1mm}
\end{equation}

Therefore, the probability that UE $v$ who has cached file $f$ is selected by requester $u$ can be written as
\vspace{-1mm}
\begin{equation}
\begin{array}{l}
P_f^{{\rm{C}}} = \mathop {\lim }\limits_{U \to \infty } \sum\limits_{k = 0}^{U - 2} {\frac{1}{{k + 1}} \cdot P_f^{(k)}} \\
= \mathop {\lim }\limits_{U \to \infty } \sum\limits_{k = 0}^{U - 2} {\frac{1}{{k + 1}} \cdot \frac{{{{({\eta _f}\psi )}^k}}}{{k!}} \cdot {e^{ - {\eta _f}\psi }} = \frac{1}{{{\eta _f}\psi }}} (1 - {e^{ - {\eta _f}\psi }})
\end{array}
\vspace{-1mm}
\end{equation}

The probability that UE $v$ meets any UE who has not cached file $f$ is $1 - {\eta _f}$, so the average number of UEs requesting file $f$ encountered by UE $v$
is $(U - 1) (1 - {\eta _f})\rho{q_f}$. Thus the average number of requesters connected to UEs who have cached   $f$ is

\vspace{-1mm}
\begin{equation}
\begin{aligned}
N_f& = \mathop {\lim }\limits_{U \to \infty } (U - 1)(1 - {\eta _f}){q_f}\rho P_f^{{\rm{C}}}\\
&= \frac{{1 - {\eta _f}}}{{{\eta _f}}} \cdot {q_f}(1 - {e^{ - {\eta _f}\psi }})
\end{aligned}
\vspace{-2mm}
\end{equation}

\subsection{Equilibrium Analysis of Evolutionary Game}\label{section:game:equilibrium-analysis}


As we consider a mobile edge network consisting of a large number of UEs, the strategy of a single UE has a negligible impact on the whole network.
Evolutionary game is an effective theoretical tool for analyzing the behavior of a large-scale group, which is often used to predict the group's selection process and finally find a dynamic equilibrium.
Therefore, we model the decision-making process of mobile UEs as an evolutionary game.

In order to achieve the final equilibrium of the evolutionary game, we present an evolutionary-game iterative algorithm as shown in the following table. The algorithm involves many rounds of interaction between UEs, where in each round, UEs change their state according to the strategies of other UEs in the previous round.
When UEs' strategies do not change anymore, the algorithm reaches an equilibrium.
Formally, we summarize the key results in the following lemmas.\footnote{Due to space limit, we leave the proof in  online technical report \cite{online}.
}

 \begin{algorithm}
	\caption{Evolutionary-Game Iterative Algorithm}
	\LinesNumbered 
	Initialize the   caching strategy ${X^{(0)}} = {\{ x_i^{(0)}\} _{U \times 1}}{\rm{ = }}0$\;
	Let ${X^\dag } = {\{ x_i^\dag \} _{U \times 1}} = {X^{(0)}}$\;
		\While{\emph{(1)}}{
			Calculate ${\eta _f}$, ${P_f}$, $ {Y_f}$\;
			\For{ i= \emph{1} \emph{to} U}{
				Calculate $x_i^* = \arg \mathop {\max }\limits_{{x_i}} {U_i}(x_i^{},x_{ - i}^\dag)$;
			}
		 ${X^*} = {\{ x_i^* \} _{U \times 1}}$
		
		${X^*} = \gamma  \cdot {X^\dag }{\rm{ +  (1 - }}\gamma ) \cdot {X^*}$
		
	\eIf{$\left\| {{X^*} - {X^\dag }} \right\| \le \varepsilon $}{
		break\;
	}{
		${X^\dag } = {X^*}$\;
	}
}
\end{algorithm}

\begin{lemma}
There exist proper values of factor $\gamma$ that guarantee algorithm 1 to converge.
\end{lemma}


\begin{lemma}
If algorithm 1 converges, it must converge to an equilibrium point of the game.
\end{lemma}



\begin{figure*}
 	\centering
	\hspace{-5mm}
	\subfigure[]{
	\includegraphics[width=2.3in,height=1.9in]{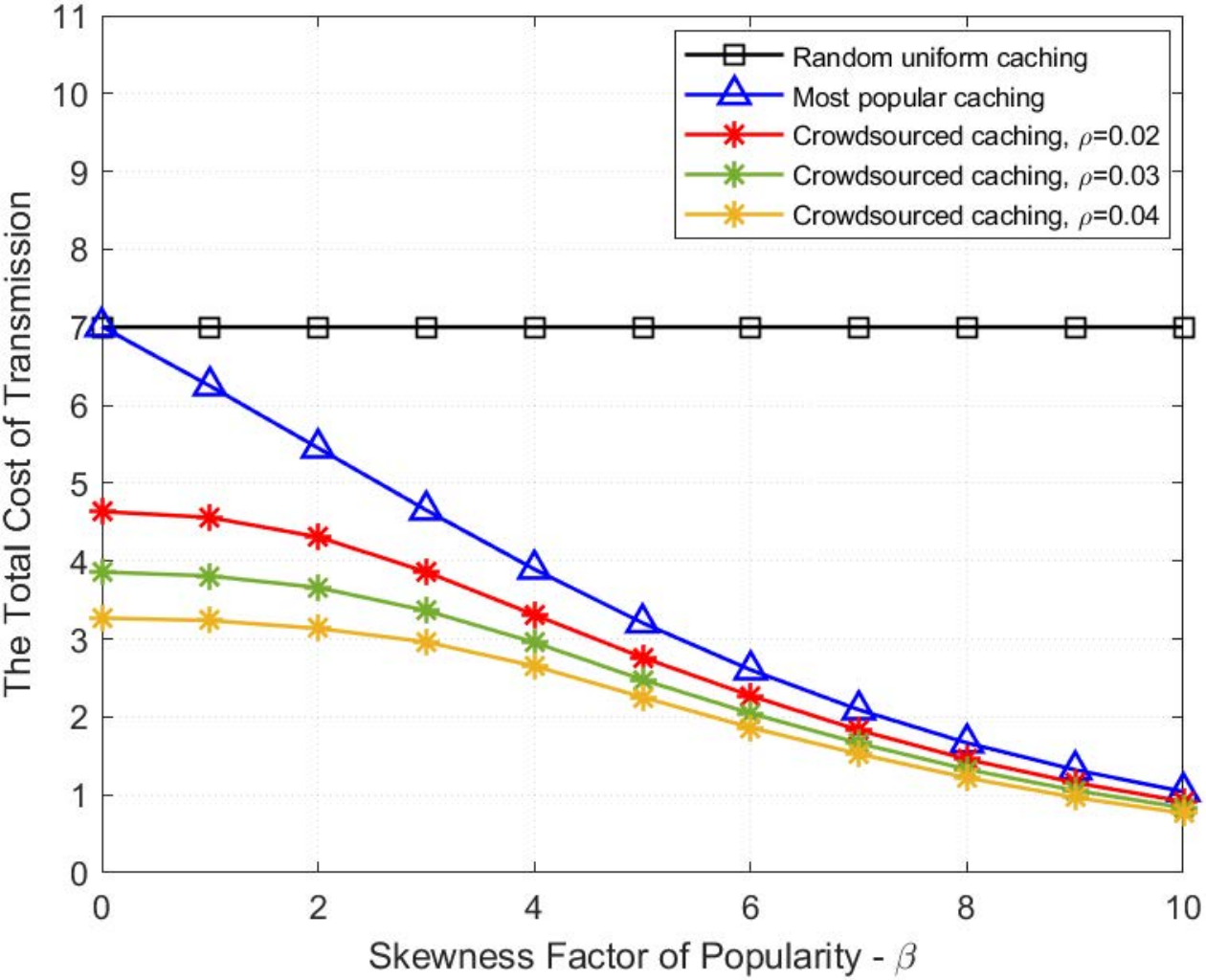}
	}
	\subfigure[]{
	\includegraphics[width=2.3in,height=1.9in]{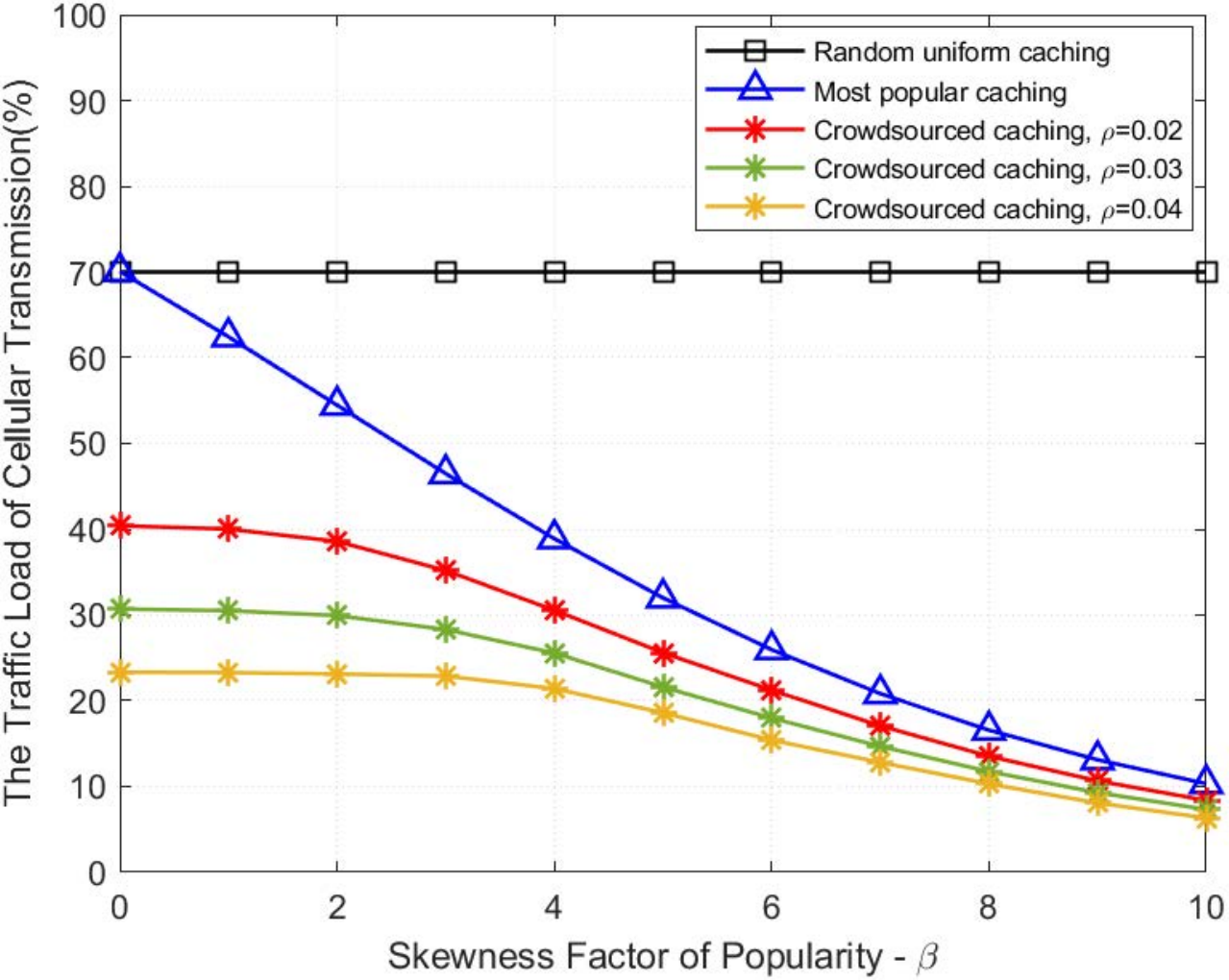}
	}
	\subfigure[]{
	\includegraphics[width=2.3in,height=1.9in]{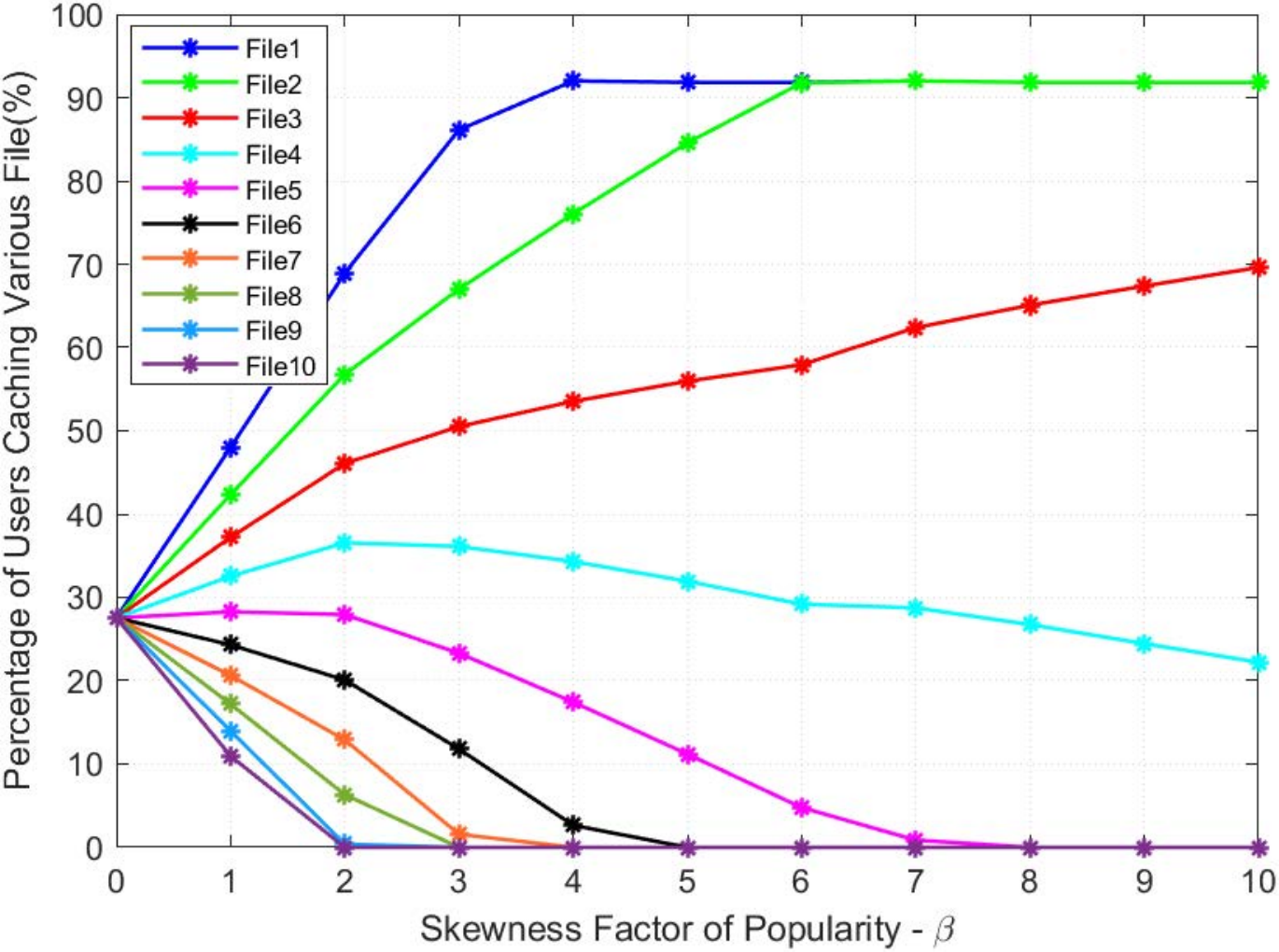}}
	\vspace{-3mm}
	\caption{(a) Total Cost of Transmission; (b)Cellular Traffic Load; (c)Percentage of Users Caching Various File vs $\beta$.}
	\label{fig:mcs-simu1}
\end{figure*}

\begin{figure*}
	\vspace{-3mm}
	\centering
	\hspace{-5mm}
	\subfigure[]{
	\includegraphics[width=2.3in,height=1.9in]{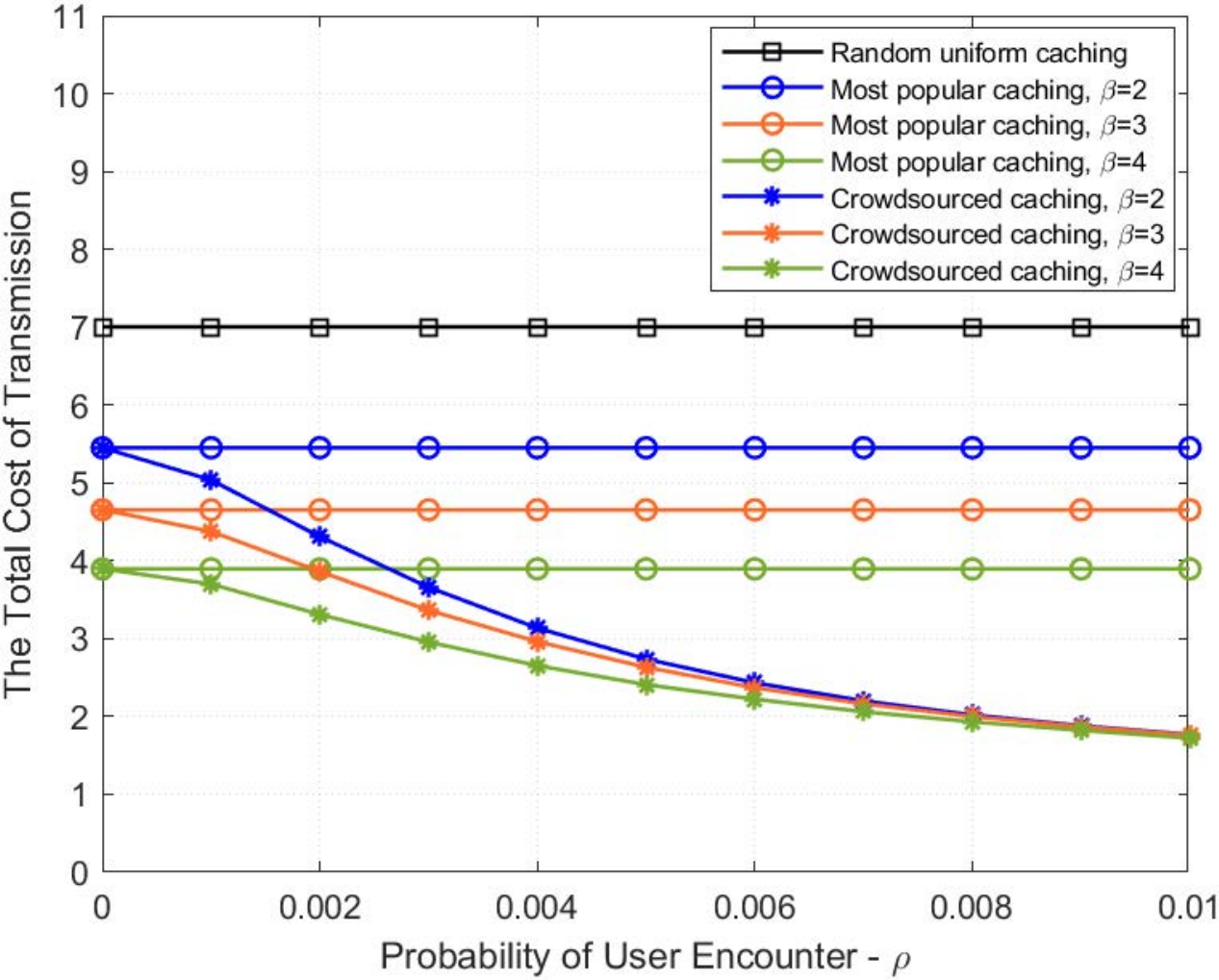}
	}
	\subfigure[]{
	\includegraphics[width=2.3in,height=1.9in]{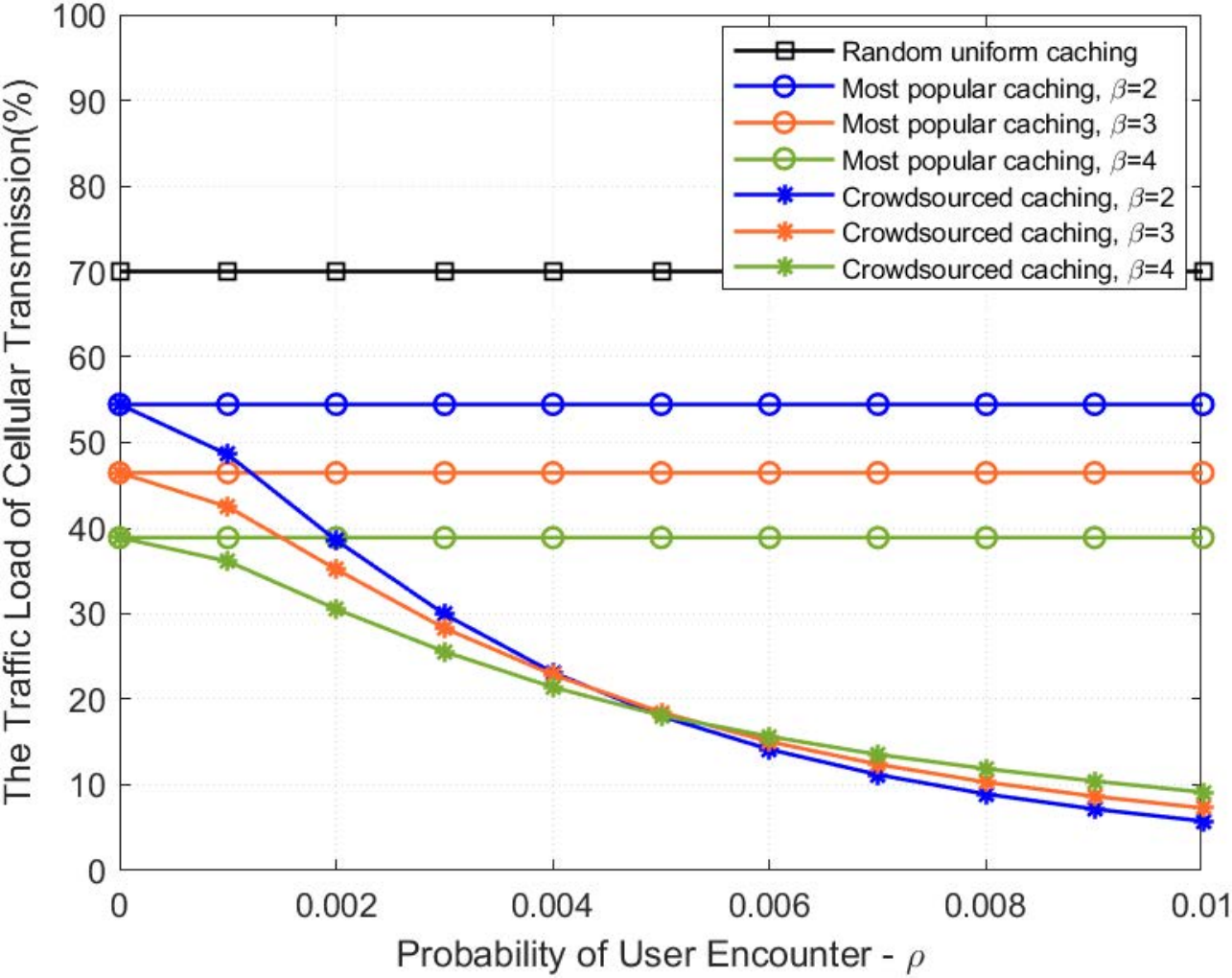}
	}
	\subfigure[]{
	\includegraphics[width=2.3in,height=1.9in]{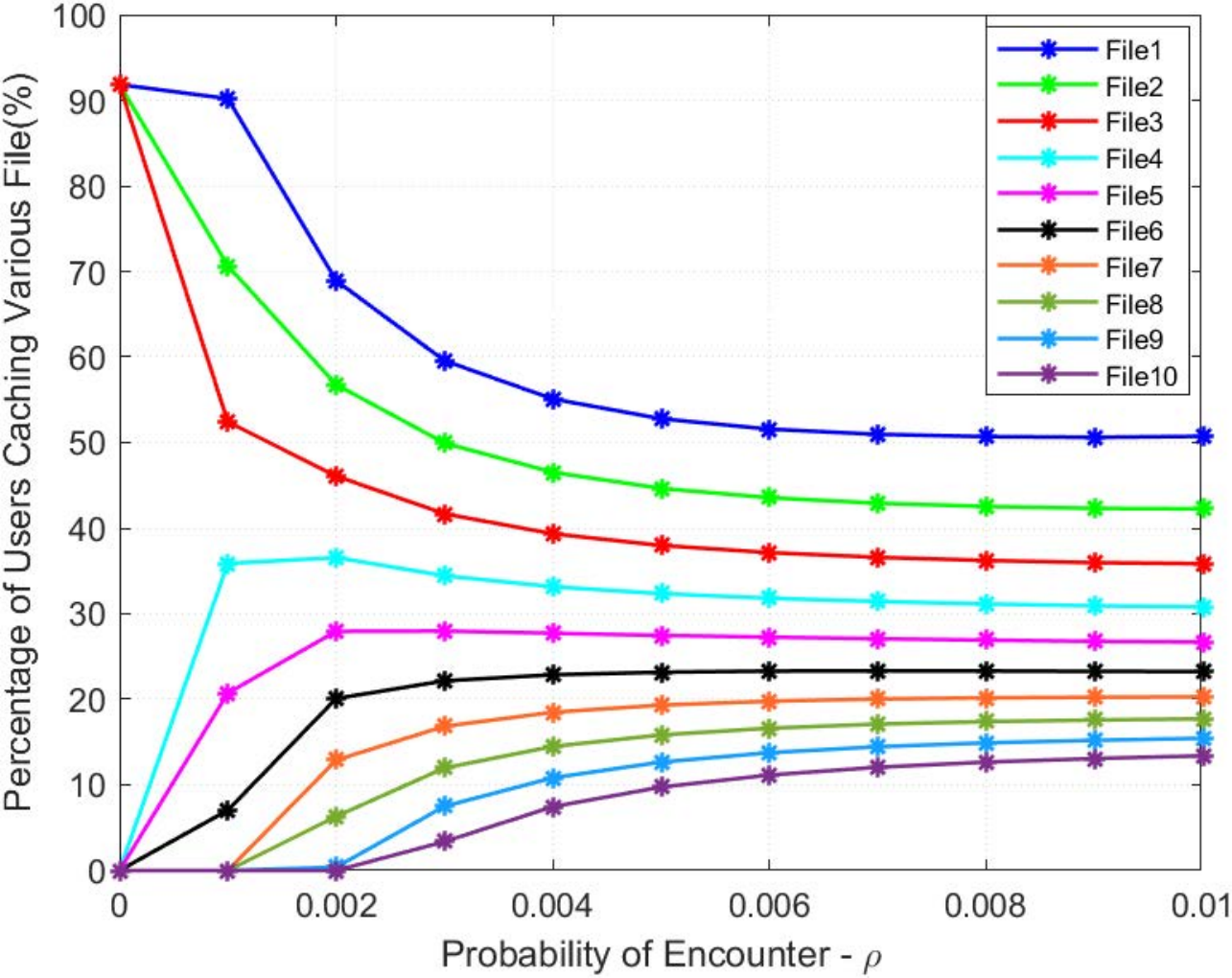}}
	\vspace{-3mm}
	\caption{(a) Total Cost of Transmission; (b)Cellular Traffic Load; (c)Percentage of Users Caching Various File vs $\rho $.}
	\label{fig:mcs-simu2}
	\vspace{-3mm}
\end{figure*}

\section{Simulations}\label{section:simulation}


We perform numerical simulations to illustrate the performance of crowdsourced caching based on the evolutionary game. We use the following parameters of scene in the simulation:
 (i) $U=1000$, $F=10$, ${s_f}=1(\forall f \in {{\cal F}})$, ${c_u}=3(\forall u \in{{\cal U}})$; (ii) the ranking of file's popularity  decreases by index in increasing order; (iii) unit cost of transmission ${\omega _L}=0.01$, ${\omega _D}=2$, ${\omega _B}=10$, reward for unit  D2D transfer $r=1.5$; (iv)as the transmission environment is relatively safe, $\theta=0.1$; (v) for limitations on energy of D2D transmission, $E = 75$, $e = 0.7$; (vi) the iteration factor is set to 0.98.

 We compare the crowdsourced caching scheme proposed in this work with the following two baseline schemes: (i) \emph{Most Popular Caching (MPC)}, where all UEs cache the most popular files; (ii) \emph{Random Uniform Caching (RUC)}, where UEs cache files randomly to fill up their storage.
 To evaluate these schemes, we use the following two performance metrics: (i) \emph{User's average cost of transmission}, denoting the average sum of cost to satisfy UE's request through local, D2D, and cellular transmission; (ii) \emph{Average traffic load of cellular transmission}, denoting the average load of UEs' requests satisfied via cellular links. In addition, the impact of various parameters on the percentage of users caching different files can also reflect the performance of the system.


Subfigure (a) of Fig.\ref{fig:mcs-simu1} and  Fig.\ref{fig:mcs-simu2} show the total transmission costs under different   $\beta$ and   $\rho$.
We can see that as $\beta $ and $\rho$ increase, the total cost of both the crowdsourced schemes and the MPC scheme continues to decrease. 
Moreover, with the increase of $\beta$, user requests are gradually concentrated on a few of the most popular files, and the probability of satisfying requests through D2D sharing increases, so the performance of the MPC schenme and our scheme are gradually approaching.
Comparing with the MPC scheme, our scheme can reduce the total cost by 42.5\%, and comparing with the RUC scheme, it can reduce the total cost   by 55.2\%.

Subfigure (b) of Fig.\ref{fig:mcs-simu1} and  Fig.\ref{fig:mcs-simu2} show the cellular transmission loads under different $\beta$ and $\rho$. We can see that as $\beta $ and $\rho$ increase, the total cost of the crowdsourced scheme continues to decrease.
With the increase of $\beta $, the slope of cellular load changing with $\rho$ gradually decreases.
Comparing with the MPC scheme, our scheme can reduce the cellular load by 21.5\%, and comparing with the RUC scheme, it can reduce the cellular load by 56.4\%.

Subfigure (c) of Fig.\ref{fig:mcs-simu1} and Fig.\ref{fig:mcs-simu2} show the percentage of UEs caching different files under different $\beta$ and $\rho$.
We can see that when the popularity of files is evenly distributed, the probability of different files being cached is all the same. With the concentration of requests, the caching probability gradually becomes scattered.
When the encounter probability is 0, UEs only cache the most popular files to meet their own needs. 

\section{Conclusion}\label{section:conclusion}



In this work, we studied a crowdsourced caching
framework for mobile edge caching, which enables mobile devices
to cache and share popular contents with each other via D2D.
We adopted 
the evolutionary game theory to analyze the UEs' content caching
and sharing strategies. 
Simulation results show that our proposed crowdsourced
caching scheme outperforms the existing schemes in terms of
the total transmission cost and the cellular load.


\end{document}